\documentclass[prl,aps,showpacs,twocolumn,floatfix]{revtex4-1}
\usepackage{amsmath,amsfonts,amssymb,bm,graphicx}
\newcommand{\bS}{{\bm S}}
\newcommand{\bk}{{\bm k}}
\newcommand{\bp}{{\bm p}}
\newcommand{\bsig}{{\bm\sigma}}
\newcommand{\e}{{\rm e}}
\newcommand{\ii}{{\rm i}}

\newcommand{\cd}{c^\dag}
\newcommand{\fd}{f^\dag}

\newcommand{\ua}{{\uparrow}}
\newcommand{\da}{{\downarrow}}
\newcommand{\vacf}{\vert0_f\rangle}

\begin{document}
\title{Orbital order and Hund's rule frustration in Kondo lattices}
\author{L. Isaev$^1$}
\author{K. Aoyama$^{2,1,3}$}
\author{I. Paul$^4$}
\author{I. Vekhter$^1$}
\affiliation{$^1$Department of Physics and Astronomy, Louisiana State
                 University, Baton Rouge, LA 80703, USA \\
	     $^2$The Hakubi Center for Advanced Research, Kyoto University,
		 Kyoto 606-8501, Japan \\
             $^3$Department of Physics, Kyoto University,
	         Kyoto 606-8502, Japan \\
	     $^4$Laboratoire Mat\'{e}riaux et Ph\'{e}nom\`{e}nes Quantiques,
	         Universit\'{e} Paris Diderot-Paris 7 \& CNRS, UMR 7162, 75205
	         Paris, France}
\begin{abstract}

 We analyze a microscopic origin of the Kondo effect-assisted orbital order in
 heavy-fermion materials. By studying the periodic two-orbital Anderson model
 with two local electrons, we show that frustration of Hund's rule coupling due
 to the Kondo effect leads to an incommensurate spiral orbital and magnetic
 order, which exists only inside the Kondo screened (heavy-electron) phase.
 This spiral state can be observed in neutron and resonant X-ray scattering
 measurements in ${\rm U}$- and ${\rm Pr}$-based heavy-fermion compounds, and
 realized in cold atomic gases, e.g. fermionic ${}^{173}{\rm Yb}$.

\end{abstract}
\maketitle

\paragraph*{Introduction.}

The dichotomy between localized and itinerant behavior of electrons in solids
often leads to a rich variety of quantum states of matter with fascinating
physical properties. In some materials $4f$ or $5f$-electrons physically move
on and off the ionic site. Even when such valence fluctuations are suppressed,
virtual transitions lead to hybridization of these electrons with the
conduction band. For one $f$-electron in a single orbital per site, the
resulting Hamiltonian describes an interplay between Kondo screening of the
local spin by conduction band (which yields large Fermi surface with heavy
quasiparticles) and local moment magnetism due to long-range
Ruderman-Kittel-Kasuya-Yosida (RKKY) interaction. This competition leads to
intriguing physics of heavy fermion (HF) metals and results in complex phase
diagrams \cite{hewson-1997} including superconducting and magnetically ordered
phases \cite{coleman-2007}.

Materials where localized electrons occupy several (nearly) degenerate atomic
orbitals display an even richer physics because the orbital degree of freedom
becomes an active participant in establishing the Kondo screened phase
\cite{hewson-1997,cox-1999}. If different orbital configurations are not
related by the time reversal symmetry, the ground state (GS) of an $f$-electron
ion may have a finite electric multipole moment \cite{hotta-2006}. The phases
associated with long-range ordering of these multipoles were observed in
various compounds, e.g. antiferroquadrupole states in ${\rm CeB_6}$
\cite{effantin-1985,erkelens-1987}, ${\rm PrPb_3}$ \cite{onimaru-2005} and
${\rm PrIr_2Zn_{20}}$ \cite{onimaru-2011}, or octupole order in ${\rm NpO_2}$
\cite{paixao-2002} and ${\rm Ce_{0.7}La_{0.3}B_6}$ \cite{mannix-2005}. As a
result orbital physics in $f$-electron materials received much recent
theoretical attention \cite{kuramoto-2009,hotta-2012}. It was also suggested
that orbital fluctuations provide a glue for unconventional superconductivity
\cite{hotta-2009,goremychkin-2004,matsubayashi-2012}, and are responsible for
the "hidden" order in ${\rm URu_2Si_2}$
\cite{chandra-2002,ressouche-2012,das-2012}.

Conventional microscopic mechanisms for the orbital order in $f$-electron
systems include an RKKY-like exchange between the multipoles mediated by
conduction electrons \cite{shiina-1997,onimaru-2005} and a direct
Heisenberg-like multipole interaction arising in the strong-coupling ($t$-$J$
like) limit of a purely $f$-electron model without the conduction band
\cite{kubo-2005}. In these cases the Kondo screening and multipole order are
antagonistic towards each other. Here we show that under certain conditions a
long-range orbital order in $f$-electron materials may exist {\it due to} the
Kondo effect.

The low-energy electronic configuration of an $f$-electron ion in the lattice
is determined by a hierarchy of energy scales (the $j-j$ coupling scheme)
\cite{hotta-2003}. First, the atomic spin-orbit interaction and crystal
electric field (CEF) splitting determine the GS multiplet in accordance with
the point symmetry double group \cite{bir-1974}. The remaining degeneracy is
partially lifted by the Hund's rule interaction. For an isolated multiorbital
Kondo impurity, the Hund coupling suppresses the Kondo temperature $T_K$
\cite{okada-1973,georges-2013}. Conversely, formation of the Kondo resonance
aims at restoring the orbital degeneracy thus {\it frustrating} the Hund
interaction.

In this Letter we use the above intuition to show that the competition between
Hund coupling and Kondo screening can give rise to combined orbital and
magnetic (generally incommensurate spiral) orders in Kondo lattices (KLs). We
consider a single-channel two-orbital periodic Anderson model with two local
electrons (the $f^2$ configuration) in the Kondo regime. In the absence of
Hund's splitting, $J$, the mixing of high- and low-spin states of an
$f$-electron ion due to their hybridization with the conduction band yields an
emergent $SO(4)$ symmetry of the problem which involves spin and orbital
$f$-electron degrees of freedom on an equal footing, and results in a
macroscopic degeneracy of the Kondo screened GS. For a finite $J$, this
symmetry is broken down to $SU(2)$ and the GS degeneracy is partially lifted.
Quantum fluctuations {\it inside the HF liquid} lead to an effective RKKY-like
interaction between magnetic and orbital degrees of freedom, which stabilizes a
long-range orbital order. In contrast to the previous works, this order cannot
exist away from the Kondo regime.

Our results are directly applicable to ${\rm U}$- and ${\rm Pr}$-based HFs in
which tetravalent ${\rm U}^{4+}$ and ${\rm Pr}^{4+}$ ($5f^2$ and $4f^2$
configurations respectively) ions have a $\Gamma_8$-type CEF GS, and the Hund
interaction is small compared to the CEF splitting. They are also relevant for
ultracold fermion gases in optical lattices, especially in light of recent
proposals to realize KL models with either special optical superlattice
structures \cite{paredes-2005}, or alkaline atoms \cite{gorshkov-2010} (e.g.
${}^{173}{\rm Yb}$).

\paragraph*{Emergent $SO(4)$ structure in the two-orbital KL model.}

We derive the KL model from a two-orbital Anderson model with two localized
electrons. First, let us consider a two-orbital Anderson {\it impurity} model
with a single conduction channel \cite{parmenter-1973}
\begin{align}
 H=&\sum_{\bk\sigma}\varepsilon_\bk\cd_{\bk\sigma}c_{\bk\sigma}+
 \frac{1}{\sqrt{N}}\sum_{\bp\sigma a}\bigl(v_{\bp a}\cd_{\bp\sigma}
 f_{a\sigma}+{\rm h.c.}\bigr)+ \nonumber \\
 &+(J-\epsilon_f)N_f-J\bigl(\bS_f^2+N_f^2/4\bigr)+UN_f(N_f-1)/2, \nonumber
\end{align}
which describes a system of conduction electrons $c_{\bk\sigma}$ (with momentum
$\bk$, spin $\sigma=\lbrace\ua,\da\rbrace$, and band dispersion
$\varepsilon_\bk$) hybridized with electrons created in two impurity orbitals
by $\fd_{a\sigma}$ ($a=1,2$) via the orbital-dependent amplitude $v_{\bp a}$.
The two orbitals correspond to a CEF GS multiplet with the binding energy
$-\epsilon_f<0$. In the above expression, $N$ is the number of lattice sites,
$N_f=\sum_{a\sigma}\fd_{a\sigma}f_{a\sigma}$ and
$\bS_f=\frac{1}{2}\sum_a\fd_{a\alpha}\bsig_{\alpha\beta}f_{a\beta}$ ($\bsig$
are the Pauli matrices) define the electron number and spin of the impurity
respectively, and $U$ is the Coulomb repulsion between localized electrons (for
simplicity, we assume identical inter- and intra-orbital interactions).
Finally, $J\geqslant0$ is the strength of Hund's rule coupling.

\begin{figure}[t]
 \begin{center}
  \includegraphics[width=\columnwidth]{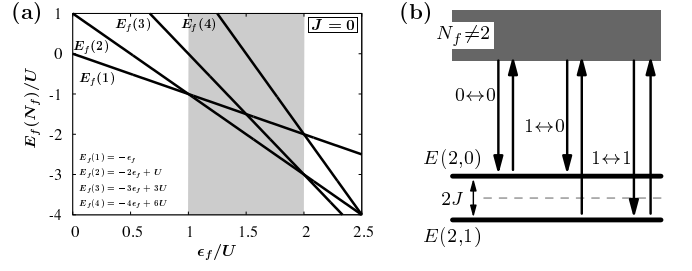}
 \end{center}
 \caption{Impurity states in the two-orbital Anderson impurity model. (a)
	  Energies $E_f(N_f)$ vs. localized level depth $\epsilon_f$ in the
	  absence of Hund's rule coupling (the $S_f$ argument can be omitted
	  when $J=0$). In the shaded region the GS has exactly $N_f=2$
	  electrons. (b) Fluctuations $S_f\leftrightarrow S^\prime_f$ in the
	  $N_f=2$ GS due to hybridization with conduction electrons which
	  contribute to the transformation \eqref{GSWT}.}
 \label{fig1}
\end{figure}

Energy levels $E_f$ for an isolated impurity with $J\ll U\sim\epsilon_f$ are
presented in Fig. \ref{fig1}(a). The GS sextet belongs to the sector $N_f=2$ if
$1\leqslant\epsilon_f/U\leqslant2$, and can be broken into subspaces with total
spins $S_f=0$ and $S_f=1$ each containing three states, as shown in Table
\ref{tab1}.

\begin{table}
 \scriptsize
 \caption{GS multiplet for an isolated impurity with $N_f=2$. $\vacf$ denotes
          a state with no local fermions.}
 \begin{center}
  \begin{tabular}{c|c}
   \hline\hline
   $S_f=0,\,\,\, E_f=-2\epsilon_f+U+J$ &
   $S_f=1,\,\,\, E_f=-2\epsilon_f+U-J$ \\
   \hline
   $\vert 00\rangle=\frac{1}{\sqrt{2}}(\fd_{1\ua}\fd_{2\da}-
    \fd_{1\da}\fd_{2\ua})\vacf$ &
   $\vert1,+1\rangle=\fd_{1\ua}\fd_{2\ua}\vacf$ \\
   $\vert s\rangle=\frac{1}{\sqrt{2}}(\fd_{1\ua}\fd_{1\da}+
    \fd_{2\ua}\fd_{2\da})\vacf$ &
   $\vert1,-1\rangle=\fd_{1\da}\fd_{2\da}\vacf$ \\
   $\vert a\rangle=\frac{1}{\sqrt{2}}(\fd_{1\ua}\fd_{1\da}-
    \fd_{2\ua}\fd_{2\da})\vacf$ &
   $\vert1,0\rangle=\frac{1}{\sqrt{2}}(\fd_{1\ua}\fd_{2\da}+
    \fd_{1\da}\fd_{2\ua})\vacf$ \\
   \hline\hline
  \end{tabular}
 \end{center}
 \label{tab1}
\end{table}

We derive the Kondo Hamiltonian via a generalized Schrieffer-Wolff
transformation \cite{muhlschlegel-1968},
$\widetilde H=\e^{\cal S}H\e^{-\cal S}$, with the generator
\begin{equation}
 {\cal S}=\frac{v}{\sqrt{N}}\sum_{\bk a\sigma\,E^\prime_fE_f}
 \biggl(\frac{\cd_{\bk\sigma}P(E^\prime_f)f_{a\sigma}P(E_f)}{E^\prime_f-E_f}-
 {\rm h.c.}\biggr),
 \label{GSWT}
\end{equation}
where $E_f$ in the sum denotes the full set of quantum numbers
$\lbrace N_fS_fS^z_f\rbrace$ corresponding to a level with energy
$E_f(N_f,S_f)$, and $P(E_f)=\vert N_fS_fS^z_f\rangle\langle N_fS_fS^z_f\vert$
is the projector on this multiplet. In writing Eq. \eqref{GSWT} we took the
hybridization $v_{\bp a}=v$ to be independent of momentum (as is usually done
in deriving the Kondo Hamiltonian \cite{hewson-1997}) and orbital index
\cite{symmetric-KLM-justification}, and assumed that
$t,\,v^2/U,\,v\ll U,\,\epsilon_f$ in order to omit the conduction electron
bandwidth (and Hund interaction) in the denominators.

The transformation \eqref{GSWT} decomposes the $N_f=2$ sextet into a doublet
$\lbrace\vert00\rangle,\,\vert s\rangle\rbrace$ and a quartet
$\lbrace\vert10\rangle,\,\vert1\pm1\rangle,\,\vert a\rangle\rbrace$. These
subspaces have different parities w.r.t. interchange of the orbitals (see Table
\ref{tab1}) and are not coupled by hybridization with the conduction band. The
resulting six-level ``Kondo'' Hamiltonian is a direct sum
$H_{\rm K}=H_4\oplus H_2$. The term $H_2\sim(1+\sigma^x)n^c_0$ has an Ising
structure ($n^c_0$ is the conduction electron density at the impurity site),
and does not involve either spin flips in the conduction channel or transitions
between impurity orbital states. Therefore it is irrelevant for the Kondo
physics.

The block $H_4$ contains coupling of the $S_f=1$ triplet {\it as well as the
singlet} $\vert a\rangle$ to the spin of conduction electrons, and can be
straightforwardly generalized to the lattice,
\begin{equation}
 H_4=\sum_{\bk\sigma}\xi_\bk\cd_{\bk\sigma}c_{\bk\sigma}+
 2J_K\sum_i{\bm\Sigma}_i{\bm s}^c_i-2J\sum_i{\bm\Sigma}_i{\bm A}_i.
 \label{so4_KLM}
\end{equation}
Here $\xi_\bk=\varepsilon_\bk-\mu_c$,
${\bm s}^c_i=\frac{1}{2}\cd_{i\alpha}\bsig_{\alpha\beta}c_{i\beta}$,
$i={\bm x}_i$ denotes lattice sites, the Kondo coupling is
$J_K=2v^2U/\bigl[(\epsilon_f-U)(2U-\epsilon_f)\bigr]>0$, chemical potential
$\mu_c$ controls the conduction band filling, and the last term describes
singlet-triplet level splitting due to Hund's interaction
\cite{GSWT-finite-J-corrections}.

The vectors ${\bm\Sigma}_i$ and ${\bm A}_i$ are spin-1/2-like objects
(${\bm\Sigma}^2_i={\bm A}^2_i=3/4$) that generate two independent (commuting)
$su(2)$ algebras. They can be expressed in terms of the on-site Hubbard
operators, $X^{M,a}_i=\vert1M\rangle\langle a\vert$,
$X^{a,M}_i=\bigl(X^{M,a}_i\bigr)^\dag$ and
$X^{M^\prime M}_i=\vert1M^\prime\rangle\langle1M\vert$, as
\begin{align}
 {\bm\Sigma}_i=(\bS_i+{\bm T}_i)/2;&\quad {\bm A}_i=(\bS_i-{\bm T}_i)/2;
 \label{so4_generators} \\
 S^+_i=\sqrt{2}\bigl(X^{0,-1}_i+X^{1,0}_i\bigr);&\quad
 T^+_i=\sqrt{2}\bigl(X^{1,a}_i-X^{a,-1}_i\bigr); \nonumber \\
 S^z_i=X^{1,1}_i-X^{-1,-1}_i;&\quad
 T^z_i=-\bigl(X^{0,a}_i+X^{a,0}_i\bigr) \nonumber
\end{align}
with $S^-_i=(S^+_i)^\dag$ and $T^-_i=(T^+_i)^\dag$. By construction $\bS_i$ has
matrix elements only within the spin-triplet sector and reduces to the $S=1$
spin operator in the limit of large $J$, while ${\bm T}_i$ contains transitions
between singlet and triplet local orbital states. Since the Hubbard operators
satisfy the constraint $X^{a,a}+\sum_MX^{M,M}=1$, and
$X^{a,a}_i={\bm T}_i^2/3-\bS^2_i/6$ and $\sum_MX^{M,M}_i=\bS^2_i/2$, the
last term in Eq. \eqref{so4_KLM} can be written up to a constant as
$2J\sum_in^a_i$, where $n^a_i$ is the occupation of the singlet state, and
hence is the Hund's energy cost. From Eq. \eqref{so4_KLM} it follows that spin
of conduction electrons couples not only to the impurity spin $\bS$ but also to
the orbital component ${\bm T}$. This interaction can be viewed as a special
spin-orbit term originating from many-body correlations.

In the following we will call a state $\vert\psi_0\rangle$ orbitally ordered if
$\langle\psi_0\vert{\bm T}_i\vert\psi_0\rangle\neq0$, and introduce the local
operators for an $f$-electron ion, $\tau^\mu_i=
\frac{1}{2}\sigma^\mu_{\alpha\beta}\fd_{i,1\alpha}f_{i,2\beta}+{\rm h.c.}$
which generate transitions between the two orbitals. One can easily verify that
{\it within the quartet subspace}, ${\bm T}_i={\bm\tau}_i$.

Operators $\bS_i$ and ${\bm T}_i$ generate an $so(4)$ algebra
\cite{kiselev-2006}, and commutativity of ${\bm\Sigma}_i$ and ${\bm A}_i$
reflects the decomposition $so(4)=su(2)\otimes su(2)$. The orbital component
${\bm T}$ is analogous to the Runge-Lenz vector in the hydrogen atom
\cite{kikoin-2001}. This hidden $so(4)$ structure of Eq. \eqref{so4_KLM} is
distinct from the explicit $SU(M)$ symmetry of the multiorbital
Coqblin-Schrieffer Hamiltonian \cite{coqblin-1969}. A model similar to Eq.
\eqref{so4_KLM} with a single impurity arises in the context of Kondo tunneling
through quantum dots \cite{eto-2000,pustilnik-2000,kikoin-2001,eto-2001}, but
to our knowledge has never been applied to KLs.

\begin{figure}[t]
 \begin{center}
  \includegraphics[width=\columnwidth]{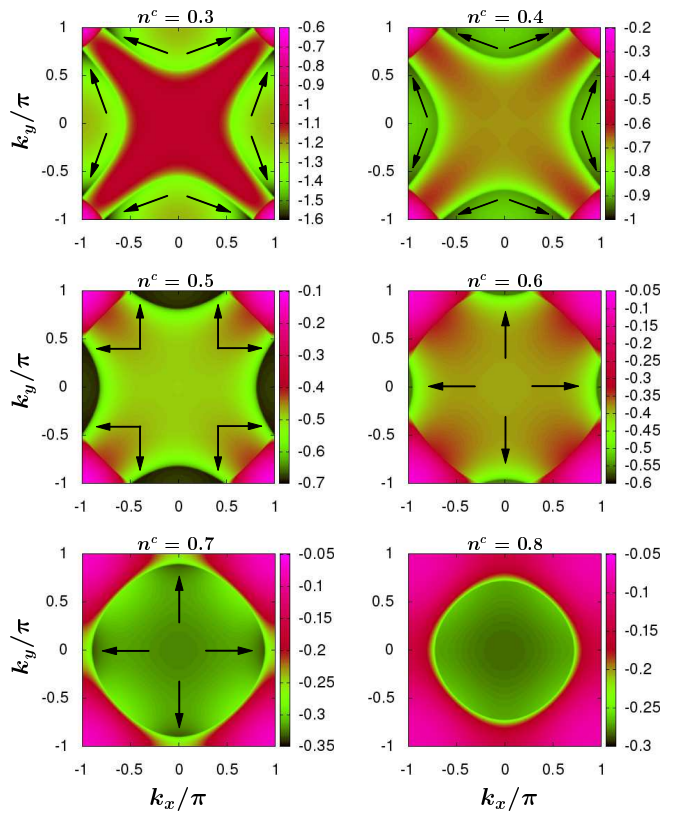}
 \end{center}
 \caption{Color maps of $tJ^{\rm RKKY}_\bk/2J^2$, Eq. \eqref{RKKY_exchange},
	  for several electron fillings $n^c$. Arrows indicate positions of the
	  minima corresponding to classical spiral states.}
 \label{fig2}
 \vspace{-0.4cm}
\end{figure}

\paragraph*{Orbitally ordered heavy-fermion state in the two-orbital KL model.}

There are two important observations about the Hamiltonian, Eq.\eqref{so4_KLM}:
(i) Despite the GS of each $f$-electron site being an $S_f=1$ triplet, the
conduction electrons are coupled to a spin-1/2 object $\bm\Sigma_i$; (ii) The
Hund's interaction is the only term that involves ${\bm A}_i$, and therefore
cannot be neglected. Eq. \eqref{so4_generators} implies that both these
operators act on physical spin and orbital degrees of freedom of the
$f$-electron ions. In the fully Kondo screened state,
$\vert{\rm HF}_{c\Sigma}\rangle$, ${\bm\Sigma}$'s form a singlet with the
conduction band, and any magnetically or orbitally ordered state corresponds to
an ordering of ${\bm A}$ pseudo-spins. While magnetic order may persist outside
of the HF regime, the orbital order exists {\it only inside} the Kondo phase.

Indeed, if we treat ${\bm\Sigma}$ and ${\bm A}$ as classical vectors, Kondo
screening does not occur, and Eq. \eqref{so4_KLM}, with $J\geqslant0$ describes
a double-exchange model \cite{anderson-1955}. The pseudo-spins ${\bm\Sigma}_i$
form a spiral whose precise shape depends on the conduction band filling, ratio
$J_K/t$, and lattice topology \cite{hamada-1995}. The vectors ${\bm A}_i$ will
follow {\it exactly the same} spiral due to the ferromagnetic Hund's coupling.
Since ${\bm\Sigma}_i$ and ${\bm A}_i$ are always locally parallel, the
situation is the same as if $J$ were infinite, i.e. at each site $S_f=1$ with
no admixture of singlet component, which is equivalent to having a pure spin-1
spiral {\it without} orbital order. This is not surprising because even for an
infinitesimal $J$ the local triplet state has a lower energy than the singlet.
The above analysis shows that a non-trivial orbital order inevitably frustrates
the (ferromagnetic) Hund term in Eq. \eqref{so4_KLM}. This {\it local}
frustration arises because quantum fluctuations associated with the Kondo
screening dynamically restore orbital degeneracy by allowing ${\bm\Sigma}_i$ to
form a singlet with the conduction band. Below we focus on the regime
$J\ll J_K$ to demonstrate how the competition between Kondo effect and Hund's
rule coupling leads to a long-range orbital order.

Returning to the quantum case, for $J=0$ the fields $\bm A_i$ decouple and the
GS of Eq. \eqref{so4_KLM} is macroscopically degenerate: $\vert\psi_0(\lbrace
A^z_i\rbrace)\rangle=\vert{\rm HF}_{c\Sigma}\rangle\otimes\vert\lbrace A^z_i
\rbrace\rangle$, where $\vert\lbrace A^z_i\rbrace\rangle$ is one of $2^N$
states which characterize the free ${\bm A}_i$ pseudo-spins. We describe the
heavy fermion state $\vert{\rm HF}_{c\Sigma}\rangle$ using the hybridization
mean-field approach (HMF) \cite{coleman-2007} with pseudo-fermion
representation
${\bm\Sigma}_i=\frac{1}{2}h^\dag_{i\alpha}\bsig_{\alpha\beta}h_{i\beta}$, and
self-consistently determine the hybridization order parameter $\chi_0=
\frac{1}{\sqrt{2}}\sum_\alpha\langle{\rm HF}_{c\Sigma}\vert\cd_{i\alpha}
h_{i\alpha}\vert{\rm HF}_{c\Sigma}\rangle$. The heavy quasiparticle dispersion
becomes $E_{\bk\tau}=\frac{1}{2}(\xi_\bk-\mu_h)+\frac{\tau}{2}R_\bk$, with
$R_\bk=\bigl[(\xi_\bk+\mu_h)^2+\frac{1}{2}(3J_K\chi_0)^2\bigr]^{1/2}$ and
$\tau=\pm1$. The $h$-fermion chemical potential, $\mu_h$, enforces the
constraint $\frac{1}{N}\sum_{i\alpha}\langle{\rm HF}_{c\Sigma}\vert
h^\dag_{i\alpha}h_{i\alpha}\vert{\rm HF}_{c\Sigma}\rangle=1$. The canonical
transformation to the quasiparticle states $\gamma_{\bk\tau\sigma}$ is given by
$c_{\bk\sigma}=\cos\frac{\rho_\bk}{2}\gamma_{\bk,+,\sigma}-\sin
\frac{\rho_\bk}{2}\gamma_{\bk,-,\sigma}$ and $h_{\bk\sigma}=\sin
\frac{\rho_\bk}{2}\gamma_{\bk,+,\sigma}+\cos\frac{\rho_\bk}{2}
\gamma_{\bk,-,\sigma}$, with $\cos\rho_\bk=(\xi_\bk+\mu_h)/R_\bk$ and
$\sin\rho_\bk=-3J_K\chi_0/\sqrt{2}R_\bk$.

When $J\ll J_K$ the Hund term in Eq. \eqref{so4_KLM} can be treated within a
second-order perturbation theory yielding an RKKY-like effective Hamiltonian
acting on the states $\vert\lbrace A^z_i\rbrace\rangle$
\begin{align}
 H_A=&P_0V(1-P_0)\bigl(E^{\rm HMF}_0-H^{\rm HMF}\bigr)^{-1}(1-P_0)VP_0=
 \nonumber \\
 &=\sum_{ij}J^{\rm RKKY}_{ij}{\bm A}_i{\bm A}_j, \nonumber
\end{align}
where $P_0$ is a projector on the degenerate HF GS manifold and
$V=-2J\sum_i{\bm\Sigma}_i\cdot{\bm A}_i$. The Fourier transform of the exchange
interaction $J^{\rm RKKY}_{ij}=\frac{1}{N}\sum_\bk\e^{\ii\bk({\bm x}_i-
{\bm x}_j)}J^{\rm RKKY}_\bk$ has the form (for conduction band filling $n^c<1$)
\begin{align}
 J^{\rm RKKY}_\bk=&\frac{2J^2}{N}\sum_\bp\cos^2\frac{\rho_{\bk+\bp}}{2}\times
 \label{RKKY_exchange} \\
 &\times\biggl[-\frac{\cos^2\frac{\rho_\bp}{2}n^\gamma_{\bp,-}}{E_{\bk+\bp,-}-
 E_{\bp,-}}+\frac{\sin^2\frac{\rho_\bp}{2}n^\gamma_{\bp,-}}{E_{\bk+\bp,-}-
 E_{\bp,+}}\biggr],
 \nonumber
\end{align}
where $n^\gamma_{\bk\tau}$ is the quasiparticle distribution function.

The semiclassical order of ${\bm A}_i$ is determined by the minima of
$J^{\rm RKKY}_\bk$. Fig. \ref{fig2} shows locations of these minima for several
electron fillings on a square lattice with nearest-neighbor electron hopping
$t$ [$\varepsilon_\bk=-2t(\cos k_x+\cos k_y)$], $N=10^6$ sites and $J_K/t=3$.
The incommensurate spiral at low filling gives way to a nearly staggered order
and finally, near $n_c=0.8$, the ordering wave vector becomes small, possibly
due to Nagaoka-like mechanism \cite{nagaoka-1966}. Since the HF phase is a
singlet ($\langle\psi_0\vert{\bm\Sigma}_i\vert\psi_0\rangle=0$),
$\langle\psi_0\vert{\bm A}_i\vert\psi_0\rangle\!=\!-\langle\psi_0\vert{\bm T}_i
\vert\psi_0\rangle\!=\!\langle\psi_0\vert{\bm S}_i\vert\psi_0\rangle$, see Eq.
\eqref{so4_generators}.

The state $\vert\psi_0\rangle$ describes an {\it orbital and real spin} spiral
of the same pitch. To the zeroth order in $J$ the occupation of
the singlet $\vert a\rangle$ and each of the triplet levels is identical, so
the orbital order manifests itself more prominently as a coherent superposition
of the triplet and singlet orbital states, rather than the difference in the
occupation numbers. For $J\neq0$ the only globally conserved quantity of the
Hamiltonian \eqref{so4_KLM} is the total spin. Hence the symmetry breaking
associated with onset of orbital order is driven by the spin sector. The
orbital order appears due to the many-body spin-orbit interaction mentioned
earlier.

\begin{figure}[t]
 \begin{center}
  \includegraphics[width=0.7\columnwidth]{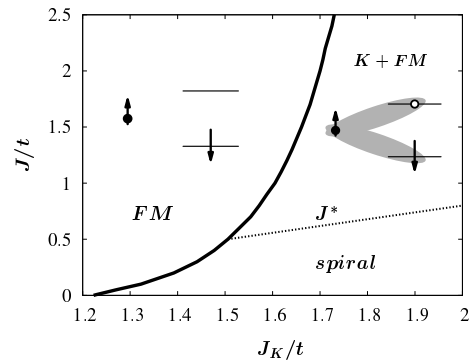}
 \end{center}
 \caption{Schematic phase diagram of the two-orbital KL model. Phases are
	  ``$FM$'' -- local moment ferromagnet, ``$K+FM$'' -- ferromagnetic
	  Kondo screened phase, ``$spiral$'' -- orbital and local-spin spiral
	  state. The solid line denotes 1st order phase transition. The dotted
	  line $J^*(J_K)$ corresponds to a transition/crossover between $K+FM$
	  and the spiral state. Shaded ellipses indicate finite Kondo
	  hybridization between conduction electrons (dark circles) and two
	  $f$-orbitals. The system size and electron density are $N=1600$ and
	  $n^c=0.6$.}
 \label{fig3}
\end{figure}

\paragraph*{Discussion.}

In multiorbital heavy-fermion materials the Kondo effect involves both spin and
orbital degrees of freedom. Our results highlight the central role of the
orbital component, that is the exotic Kondo screening of the pseudo-spin-$1/2$
object ${\bm\Sigma}_i$ which stabilizes non-trivial orbital order via a
many-body spin-orbit interaction. Formation of the local Kondo singlets
competes with Hund's rule coupling by enhancing orbital fluctuations and {\it
dynamically} restoring orbital degeneracies. This Hund's rule {\it frustration}
is the fundamental mechanism that leads to a RKKY-like exchange between
orbital degrees of freedom and drives their long-range ordering. The RKKY
interaction obtained in our work exists {\it due} to the Kondo effect, in sharp
contrast with theories of quadrupolar Kondo effect
\cite{cox-1987,shiina-1997,onimaru-2005,kubo-2005} which contend that the RKKY
coupling between electric quadrupoles {\it competes} with Kondo screening.

Our prediction of coupled orbital and magnetic orders can be tested in neutron
and resonant X-ray scattering experiments {\it inside} the heavy-electron phase
of $5f$ actinide and ${\rm Pr}$-based compounds in which atomic spin-orbit
coupling and CEF stabilize an orbitally-degenerate local GS. The pitch of the
spiral is carrier-density dependent (see Fig. \ref{fig2}) and can be tuned by
doping or pressure.

We derived the RKKY interaction \eqref{RKKY_exchange} for $J\ll J_K$ when the
pseudospin ${\bm\Sigma}$ is Kondo screened while another pseudospin ${\bm A}$
remains unscreened. For $J\gg J_K$ the singlet $\vert a\rangle$ is separated by
an energy gap from the triplet states, and Eq. \eqref{so4_KLM} reduces to the
underscreened $S=1$ KL model studied in Refs. \cite{perkins-2007,thomas-2011}.
Using a modified HMF theory \cite{coleman-2007}, it was shown that the system
exhibits a coexistence of the Kondo effect and ferromagnetism. In general, the
two regimes at small and large $J$ are separated by a quantum phase transition
because local orbitals realize an irreducible representation of the crystal
symmetry group and an orbital order would break at least this discrete
symmetry. However, in the presence of the strong spin-orbit interaction the
phase transition can become a crossover. Fig. \ref{fig3} presents a schematic
phase diagram of the Hamiltonian \eqref{so4_KLM} computed on a square lattice
using the mean-field approach of Ref. \cite{eto-2001}.

The spiral states obtained in our analysis are semiclassical. Since ${\bm A}_i$
behaves as a spin-$1/2$ object, quantum fluctuations (especially for frustrated
materials \cite{diep-2004}) may destabilize static order in favor of quantum
disordered phases. In our case quantum effects may lead to even more exotic
para-orbital states, e.g. combined orbital and spin liquids, that so far
received little attention.

\paragraph*{Acknowledgments.}

This work was supported in part by NSF Grant DMR-1105339 (L. I. and I. V.), and
started thanks to the visitor program under NSF EPSCoR Cooperative Agreement
No. EPS-1003897 with additional support from the Louisiana Board of Regents. I.
P. is grateful to E. Boulat for illuminating discussions. L. I. acknowledges
discussions with D. Solenov.


\begin{thebibliography}{43}%
\makeatletter
\providecommand \@ifxundefined [1]{%
 \@ifx{#1\undefined}
}%
\providecommand \@ifnum [1]{%
 \ifnum #1\expandafter \@firstoftwo
 \else \expandafter \@secondoftwo
 \fi
}%
\providecommand \@ifx [1]{%
 \ifx #1\expandafter \@firstoftwo
 \else \expandafter \@secondoftwo
 \fi
}%
\providecommand \natexlab [1]{#1}%
\providecommand \enquote  [1]{``#1''}%
\providecommand \bibnamefont  [1]{#1}%
\providecommand \bibfnamefont [1]{#1}%
\providecommand \citenamefont [1]{#1}%
\providecommand \href@noop [0]{\@secondoftwo}%
\providecommand \href [0]{\begingroup \@sanitize@url \@href}%
\providecommand \@href[1]{\@@startlink{#1}\@@href}%
\providecommand \@@href[1]{\endgroup#1\@@endlink}%
\providecommand \@sanitize@url [0]{\catcode `\\12\catcode `\$12\catcode
  `\&12\catcode `\#12\catcode `\^12\catcode `\_12\catcode `\%12\relax}%
\providecommand \@@startlink[1]{}%
\providecommand \@@endlink[0]{}%
\providecommand \url  [0]{\begingroup\@sanitize@url \@url }%
\providecommand \@url [1]{\endgroup\@href {#1}{\urlprefix }}%
\providecommand \urlprefix  [0]{URL }%
\providecommand \Eprint [0]{\href }%
\providecommand \doibase [0]{http://dx.doi.org/}%
\providecommand \selectlanguage [0]{\@gobble}%
\providecommand \bibinfo  [0]{\@secondoftwo}%
\providecommand \bibfield  [0]{\@secondoftwo}%
\providecommand \translation [1]{[#1]}%
\providecommand \BibitemOpen [0]{}%
\providecommand \bibitemStop [0]{}%
\providecommand \bibitemNoStop [0]{.\EOS\space}%
\providecommand \EOS [0]{\spacefactor3000\relax}%
\providecommand \BibitemShut  [1]{\csname bibitem#1\endcsname}%
\let\auto@bib@innerbib\@empty
\bibitem [{\citenamefont {Hewson}(1997)}]{hewson-1997}%
  \BibitemOpen
  \bibfield  {author} {\bibinfo {author} {\bibfnamefont {A.~C.}\ \bibnamefont
  {Hewson}},\ }\href {http://books.google.com/books?id=fPzgHneNFDAC} {\emph
  {\bibinfo {title} {The Kondo Problem to Heavy Fermions}}}\ (\bibinfo
  {publisher} {Cambridge University Press},\ \bibinfo {year}
  {1997})\BibitemShut {NoStop}%
\bibitem [{\citenamefont {Coleman}(2007)}]{coleman-2007}%
  \BibitemOpen
  \bibfield  {author} {\bibinfo {author} {\bibfnamefont {P.}~\bibnamefont
  {Coleman}},\ }\href@noop {} {\bibfield  {journal} {\bibinfo  {journal}
  {Handbook of Magnetism and Advanced Magnetic Materials, Vol. 1, H. Kronmuller
  and S. Parkin (eds.)}\ } (\bibinfo {year} {2007})}\BibitemShut {NoStop}%
\bibitem [{\citenamefont {Cox}\ and\ \citenamefont
  {Zawadowski}(1999)}]{cox-1999}%
  \BibitemOpen
  \bibfield  {author} {\bibinfo {author} {\bibfnamefont {D.~L.}\ \bibnamefont
  {Cox}}\ and\ \bibinfo {author} {\bibfnamefont {A.}~\bibnamefont
  {Zawadowski}},\ }\href {http://books.google.com/books?id=Z7BSOUUtmDAC} {\emph
  {\bibinfo {title} {Exotic Kondo Effects in Metals: Magnetic Ions in a
  Crystalline Electric Field and Tunnelling Centres}}}\ (\bibinfo  {publisher}
  {Taylor \& Francis},\ \bibinfo {year} {1999})\BibitemShut {NoStop}%
\bibitem [{\citenamefont {Hotta}(2006)}]{hotta-2006}%
  \BibitemOpen
  \bibfield  {author} {\bibinfo {author} {\bibfnamefont {T.}~\bibnamefont
  {Hotta}},\ }\href {http://stacks.iop.org/0034-4885/69/i=7/a=R02} {\bibfield
  {journal} {\bibinfo  {journal} {Reports on Progress in Physics}\ }\textbf
  {\bibinfo {volume} {69}},\ \bibinfo {pages} {2061} (\bibinfo {year}
  {2006})}\BibitemShut {NoStop}%
\bibitem [{\citenamefont {Effantin}\ \emph {et~al.}(1985)\citenamefont
  {Effantin}, \citenamefont {Rossat-Mignod}, \citenamefont {Burlet},
  \citenamefont {Bartholin}, \citenamefont {Kunii},\ and\ \citenamefont
  {Kasuya}}]{effantin-1985}%
  \BibitemOpen
  \bibfield  {author} {\bibinfo {author} {\bibfnamefont {J.}~\bibnamefont
  {Effantin}}, \bibinfo {author} {\bibfnamefont {J.}~\bibnamefont
  {Rossat-Mignod}}, \bibinfo {author} {\bibfnamefont {P.}~\bibnamefont
  {Burlet}}, \bibinfo {author} {\bibfnamefont {H.}~\bibnamefont {Bartholin}},
  \bibinfo {author} {\bibfnamefont {S.}~\bibnamefont {Kunii}}, \ and\ \bibinfo
  {author} {\bibfnamefont {T.}~\bibnamefont {Kasuya}},\ }\href {\doibase
  10.1016/0304-8853(85)90382-8} {\bibfield  {journal} {\bibinfo  {journal}
  {Journal of Magnetism and Magnetic Materials}\ }\textbf {\bibinfo {volume}
  {47 -- 48}},\ \bibinfo {pages} {145 } (\bibinfo {year} {1985})}\BibitemShut
  {NoStop}%
\bibitem [{\citenamefont {Erkelens}\ \emph {et~al.}(1987)\citenamefont
  {Erkelens}, \citenamefont {Regnault}, \citenamefont {Burlet}, \citenamefont
  {Rossat-Mignod}, \citenamefont {Kunii},\ and\ \citenamefont
  {Kasuya}}]{erkelens-1987}%
  \BibitemOpen
  \bibfield  {author} {\bibinfo {author} {\bibfnamefont {W.}~\bibnamefont
  {Erkelens}}, \bibinfo {author} {\bibfnamefont {L.}~\bibnamefont {Regnault}},
  \bibinfo {author} {\bibfnamefont {P.}~\bibnamefont {Burlet}}, \bibinfo
  {author} {\bibfnamefont {J.}~\bibnamefont {Rossat-Mignod}}, \bibinfo {author}
  {\bibfnamefont {S.}~\bibnamefont {Kunii}}, \ and\ \bibinfo {author}
  {\bibfnamefont {T.}~\bibnamefont {Kasuya}},\ }\href {\doibase
  10.1016/0304-8853(87)90522-1} {\bibfield  {journal} {\bibinfo  {journal}
  {Journal of Magnetism and Magnetic Materials}\ }\textbf {\bibinfo {volume}
  {63 -- 64}},\ \bibinfo {pages} {61 } (\bibinfo {year} {1987})}\BibitemShut
  {NoStop}%
\bibitem [{\citenamefont {Onimaru}\ \emph {et~al.}(2005)\citenamefont
  {Onimaru}, \citenamefont {Sakakibara}, \citenamefont {Aso}, \citenamefont
  {Yoshizawa}, \citenamefont {Suzuki},\ and\ \citenamefont
  {Takeuchi}}]{onimaru-2005}%
  \BibitemOpen
  \bibfield  {author} {\bibinfo {author} {\bibfnamefont {T.}~\bibnamefont
  {Onimaru}}, \bibinfo {author} {\bibfnamefont {T.}~\bibnamefont {Sakakibara}},
  \bibinfo {author} {\bibfnamefont {N.}~\bibnamefont {Aso}}, \bibinfo {author}
  {\bibfnamefont {H.}~\bibnamefont {Yoshizawa}}, \bibinfo {author}
  {\bibfnamefont {H.~S.}\ \bibnamefont {Suzuki}}, \ and\ \bibinfo {author}
  {\bibfnamefont {T.}~\bibnamefont {Takeuchi}},\ }\href {\doibase
  10.1103/PhysRevLett.94.197201} {\bibfield  {journal} {\bibinfo  {journal}
  {Phys. Rev. Lett.}\ }\textbf {\bibinfo {volume} {94}},\ \bibinfo {pages}
  {197201} (\bibinfo {year} {2005})}\BibitemShut {NoStop}%
\bibitem [{\citenamefont {Onimaru}\ \emph {et~al.}(2011)\citenamefont
  {Onimaru}, \citenamefont {Matsumoto}, \citenamefont {Inoue}, \citenamefont
  {Umeo}, \citenamefont {Sakakibara}, \citenamefont {Karaki}, \citenamefont
  {Kubota},\ and\ \citenamefont {Takabatake}}]{onimaru-2011}%
  \BibitemOpen
  \bibfield  {author} {\bibinfo {author} {\bibfnamefont {T.}~\bibnamefont
  {Onimaru}}, \bibinfo {author} {\bibfnamefont {K.~T.}\ \bibnamefont
  {Matsumoto}}, \bibinfo {author} {\bibfnamefont {Y.~F.}\ \bibnamefont
  {Inoue}}, \bibinfo {author} {\bibfnamefont {K.}~\bibnamefont {Umeo}},
  \bibinfo {author} {\bibfnamefont {T.}~\bibnamefont {Sakakibara}}, \bibinfo
  {author} {\bibfnamefont {Y.}~\bibnamefont {Karaki}}, \bibinfo {author}
  {\bibfnamefont {M.}~\bibnamefont {Kubota}}, \ and\ \bibinfo {author}
  {\bibfnamefont {T.}~\bibnamefont {Takabatake}},\ }\href {\doibase
  10.1103/PhysRevLett.106.177001} {\bibfield  {journal} {\bibinfo  {journal}
  {Phys. Rev. Lett.}\ }\textbf {\bibinfo {volume} {106}},\ \bibinfo {pages}
  {177001} (\bibinfo {year} {2011})}\BibitemShut {NoStop}%
\bibitem [{\citenamefont {Paix\~ao}\ \emph {et~al.}(2002)\citenamefont
  {Paix\~ao}, \citenamefont {Detlefs}, \citenamefont {Longfield}, \citenamefont
  {Caciuffo}, \citenamefont {Santini}, \citenamefont {Bernhoeft}, \citenamefont
  {Rebizant},\ and\ \citenamefont {Lander}}]{paixao-2002}%
  \BibitemOpen
  \bibfield  {author} {\bibinfo {author} {\bibfnamefont {J.~A.}\ \bibnamefont
  {Paix\~ao}}, \bibinfo {author} {\bibfnamefont {C.}~\bibnamefont {Detlefs}},
  \bibinfo {author} {\bibfnamefont {M.~J.}\ \bibnamefont {Longfield}}, \bibinfo
  {author} {\bibfnamefont {R.}~\bibnamefont {Caciuffo}}, \bibinfo {author}
  {\bibfnamefont {P.}~\bibnamefont {Santini}}, \bibinfo {author} {\bibfnamefont
  {N.}~\bibnamefont {Bernhoeft}}, \bibinfo {author} {\bibfnamefont
  {J.}~\bibnamefont {Rebizant}}, \ and\ \bibinfo {author} {\bibfnamefont
  {G.~H.}\ \bibnamefont {Lander}},\ }\href {\doibase
  10.1103/PhysRevLett.89.187202} {\bibfield  {journal} {\bibinfo  {journal}
  {Phys. Rev. Lett.}\ }\textbf {\bibinfo {volume} {89}},\ \bibinfo {pages}
  {187202} (\bibinfo {year} {2002})}\BibitemShut {NoStop}%
\bibitem [{\citenamefont {Mannix}\ \emph {et~al.}(2005)\citenamefont {Mannix},
  \citenamefont {Tanaka}, \citenamefont {Carbone}, \citenamefont {Bernhoeft},\
  and\ \citenamefont {Kunii}}]{mannix-2005}%
  \BibitemOpen
  \bibfield  {author} {\bibinfo {author} {\bibfnamefont {D.}~\bibnamefont
  {Mannix}}, \bibinfo {author} {\bibfnamefont {Y.}~\bibnamefont {Tanaka}},
  \bibinfo {author} {\bibfnamefont {D.}~\bibnamefont {Carbone}}, \bibinfo
  {author} {\bibfnamefont {N.}~\bibnamefont {Bernhoeft}}, \ and\ \bibinfo
  {author} {\bibfnamefont {S.}~\bibnamefont {Kunii}},\ }\href {\doibase
  10.1103/PhysRevLett.95.117206} {\bibfield  {journal} {\bibinfo  {journal}
  {Phys. Rev. Lett.}\ }\textbf {\bibinfo {volume} {95}},\ \bibinfo {pages}
  {117206} (\bibinfo {year} {2005})}\BibitemShut {NoStop}%
\bibitem [{\citenamefont {Kuramoto}\ \emph {et~al.}(2009)\citenamefont
  {Kuramoto}, \citenamefont {Kusunose},\ and\ \citenamefont
  {Kiss}}]{kuramoto-2009}%
  \BibitemOpen
  \bibfield  {author} {\bibinfo {author} {\bibfnamefont {Y.}~\bibnamefont
  {Kuramoto}}, \bibinfo {author} {\bibfnamefont {H.}~\bibnamefont {Kusunose}},
  \ and\ \bibinfo {author} {\bibfnamefont {A.}~\bibnamefont {Kiss}},\ }\href
  {\doibase 10.1143/JPSJ.78.072001} {\bibfield  {journal} {\bibinfo  {journal}
  {Journal of the Physical Society of Japan}\ }\textbf {\bibinfo {volume}
  {78}},\ \bibinfo {pages} {072001} (\bibinfo {year} {2009})}\BibitemShut
  {NoStop}%
\bibitem [{\citenamefont {Hotta}(2012)}]{hotta-2012}%
  \BibitemOpen
  \bibfield  {author} {\bibinfo {author} {\bibfnamefont {T.}~\bibnamefont
  {Hotta}},\ }\href {\doibase 10.1155/2012/762798} {\bibfield  {journal}
  {\bibinfo  {journal} {Physics Research International}\ }\textbf {\bibinfo
  {volume} {2012}},\ \bibinfo {pages} {762798} (\bibinfo {year}
  {2012})}\BibitemShut {NoStop}%
\bibitem [{\citenamefont {Hotta}(2009)}]{hotta-2009}%
  \BibitemOpen
  \bibfield  {author} {\bibinfo {author} {\bibfnamefont {T.}~\bibnamefont
  {Hotta}},\ }\href {\doibase 10.1143/JPSJ.78.123710} {\bibfield  {journal}
  {\bibinfo  {journal} {Journal of the Physical Society of Japan}\ }\textbf
  {\bibinfo {volume} {78}},\ \bibinfo {pages} {123710} (\bibinfo {year}
  {2009})}\BibitemShut {NoStop}%
\bibitem [{\citenamefont {Goremychkin}\ \emph {et~al.}(2004)\citenamefont
  {Goremychkin}, \citenamefont {Osborn}, \citenamefont {Bauer}, \citenamefont
  {Maple}, \citenamefont {Frederick}, \citenamefont {Yuhasz}, \citenamefont
  {Woodward},\ and\ \citenamefont {Lynn}}]{goremychkin-2004}%
  \BibitemOpen
  \bibfield  {author} {\bibinfo {author} {\bibfnamefont {E.~A.}\ \bibnamefont
  {Goremychkin}}, \bibinfo {author} {\bibfnamefont {R.}~\bibnamefont {Osborn}},
  \bibinfo {author} {\bibfnamefont {E.~D.}\ \bibnamefont {Bauer}}, \bibinfo
  {author} {\bibfnamefont {M.~B.}\ \bibnamefont {Maple}}, \bibinfo {author}
  {\bibfnamefont {N.~A.}\ \bibnamefont {Frederick}}, \bibinfo {author}
  {\bibfnamefont {W.~M.}\ \bibnamefont {Yuhasz}}, \bibinfo {author}
  {\bibfnamefont {F.~M.}\ \bibnamefont {Woodward}}, \ and\ \bibinfo {author}
  {\bibfnamefont {J.~W.}\ \bibnamefont {Lynn}},\ }\href {\doibase
  10.1103/PhysRevLett.93.157003} {\bibfield  {journal} {\bibinfo  {journal}
  {Phys. Rev. Lett.}\ }\textbf {\bibinfo {volume} {93}},\ \bibinfo {pages}
  {157003} (\bibinfo {year} {2004})}\BibitemShut {NoStop}%
\bibitem [{\citenamefont {Matsubayashi}\ \emph {et~al.}(2012)\citenamefont
  {Matsubayashi}, \citenamefont {Tanaka}, \citenamefont {Sakai}, \citenamefont
  {Nakatsuji}, \citenamefont {Kubo},\ and\ \citenamefont
  {Uwatoko}}]{matsubayashi-2012}%
  \BibitemOpen
  \bibfield  {author} {\bibinfo {author} {\bibfnamefont {K.}~\bibnamefont
  {Matsubayashi}}, \bibinfo {author} {\bibfnamefont {T.}~\bibnamefont
  {Tanaka}}, \bibinfo {author} {\bibfnamefont {A.}~\bibnamefont {Sakai}},
  \bibinfo {author} {\bibfnamefont {S.}~\bibnamefont {Nakatsuji}}, \bibinfo
  {author} {\bibfnamefont {Y.}~\bibnamefont {Kubo}}, \ and\ \bibinfo {author}
  {\bibfnamefont {Y.}~\bibnamefont {Uwatoko}},\ }\href {\doibase
  10.1103/PhysRevLett.109.187004} {\bibfield  {journal} {\bibinfo  {journal}
  {Phys. Rev. Lett.}\ }\textbf {\bibinfo {volume} {109}},\ \bibinfo {pages}
  {187004} (\bibinfo {year} {2012})}\BibitemShut {NoStop}%
\bibitem [{\citenamefont {Chandra}\ \emph {et~al.}(2002)\citenamefont
  {Chandra}, \citenamefont {Coleman}, \citenamefont {Mydosh},\ and\
  \citenamefont {Tripathi}}]{chandra-2002}%
  \BibitemOpen
  \bibfield  {author} {\bibinfo {author} {\bibfnamefont {P.}~\bibnamefont
  {Chandra}}, \bibinfo {author} {\bibfnamefont {P.}~\bibnamefont {Coleman}},
  \bibinfo {author} {\bibfnamefont {J.~A.}\ \bibnamefont {Mydosh}}, \ and\
  \bibinfo {author} {\bibfnamefont {V.}~\bibnamefont {Tripathi}},\ }\href
  {\doibase 10.1038/nature00795} {\bibfield  {journal} {\bibinfo  {journal}
  {Nature}\ }\textbf {\bibinfo {volume} {417}},\ \bibinfo {pages} {831}
  (\bibinfo {year} {2002})}\BibitemShut {NoStop}%
\bibitem [{\citenamefont {Ressouche}\ \emph {et~al.}(2012)\citenamefont
  {Ressouche}, \citenamefont {Ballou}, \citenamefont {Bourdarot}, \citenamefont
  {Aoki}, \citenamefont {Simonet}, \citenamefont {Fernandez-Diaz},
  \citenamefont {Stunault},\ and\ \citenamefont {Flouquet}}]{ressouche-2012}%
  \BibitemOpen
  \bibfield  {author} {\bibinfo {author} {\bibfnamefont {E.}~\bibnamefont
  {Ressouche}}, \bibinfo {author} {\bibfnamefont {R.}~\bibnamefont {Ballou}},
  \bibinfo {author} {\bibfnamefont {F.}~\bibnamefont {Bourdarot}}, \bibinfo
  {author} {\bibfnamefont {D.}~\bibnamefont {Aoki}}, \bibinfo {author}
  {\bibfnamefont {V.}~\bibnamefont {Simonet}}, \bibinfo {author} {\bibfnamefont
  {M.~T.}\ \bibnamefont {Fernandez-Diaz}}, \bibinfo {author} {\bibfnamefont
  {A.}~\bibnamefont {Stunault}}, \ and\ \bibinfo {author} {\bibfnamefont
  {J.}~\bibnamefont {Flouquet}},\ }\href {\doibase
  10.1103/PhysRevLett.109.067202} {\bibfield  {journal} {\bibinfo  {journal}
  {Phys. Rev. Lett.}\ }\textbf {\bibinfo {volume} {109}},\ \bibinfo {pages}
  {067202} (\bibinfo {year} {2012})}\BibitemShut {NoStop}%
\bibitem [{\citenamefont {Das}(2012)}]{das-2012}%
  \BibitemOpen
  \bibfield  {author} {\bibinfo {author} {\bibfnamefont {T.}~\bibnamefont
  {Das}},\ }\href {\doibase 10.1038/srep00596} {\bibfield  {journal} {\bibinfo
  {journal} {Sci. Rep.}\ }\textbf {\bibinfo {volume} {2}},\ \bibinfo {pages}
  {596} (\bibinfo {year} {2012})}\BibitemShut {NoStop}%
\bibitem [{\citenamefont {Shiina}\ \emph {et~al.}(1997)\citenamefont {Shiina},
  \citenamefont {Shiba},\ and\ \citenamefont {Thalmeier}}]{shiina-1997}%
  \BibitemOpen
  \bibfield  {author} {\bibinfo {author} {\bibfnamefont {R.}~\bibnamefont
  {Shiina}}, \bibinfo {author} {\bibfnamefont {H.}~\bibnamefont {Shiba}}, \
  and\ \bibinfo {author} {\bibfnamefont {P.}~\bibnamefont {Thalmeier}},\ }\href
  {\doibase 10.1143/JPSJ.66.1741} {\bibfield  {journal} {\bibinfo  {journal}
  {Journal of the Physical Society of Japan}\ }\textbf {\bibinfo {volume}
  {66}},\ \bibinfo {pages} {1741} (\bibinfo {year} {1997})}\BibitemShut
  {NoStop}%
\bibitem [{\citenamefont {Kubo}\ and\ \citenamefont {Hotta}(2005)}]{kubo-2005}%
  \BibitemOpen
  \bibfield  {author} {\bibinfo {author} {\bibfnamefont {K.}~\bibnamefont
  {Kubo}}\ and\ \bibinfo {author} {\bibfnamefont {T.}~\bibnamefont {Hotta}},\
  }\href {\doibase 10.1103/PhysRevB.71.140404} {\bibfield  {journal} {\bibinfo
  {journal} {Phys. Rev. B}\ }\textbf {\bibinfo {volume} {71}},\ \bibinfo
  {pages} {140404} (\bibinfo {year} {2005})}\BibitemShut {NoStop}%
\bibitem [{\citenamefont {Hotta}\ and\ \citenamefont
  {Ueda}(2003)}]{hotta-2003}%
  \BibitemOpen
  \bibfield  {author} {\bibinfo {author} {\bibfnamefont {T.}~\bibnamefont
  {Hotta}}\ and\ \bibinfo {author} {\bibfnamefont {K.}~\bibnamefont {Ueda}},\
  }\href {\doibase 10.1103/PhysRevB.67.104518} {\bibfield  {journal} {\bibinfo
  {journal} {Phys. Rev. B}\ }\textbf {\bibinfo {volume} {67}},\ \bibinfo
  {pages} {104518} (\bibinfo {year} {2003})}\BibitemShut {NoStop}%
\bibitem [{\citenamefont {Bir}\ and\ \citenamefont {Pikus}(1974)}]{bir-1974}%
  \BibitemOpen
  \bibfield  {author} {\bibinfo {author} {\bibfnamefont {G.}~\bibnamefont
  {Bir}}\ and\ \bibinfo {author} {\bibfnamefont {G.}~\bibnamefont {Pikus}},\
  }\href {http://books.google.com/books?id=38m2QgAACAAJ} {\emph {\bibinfo
  {title} {Symmetry and Strain-Induced Effects in Semiconductors}}}\ (\bibinfo
  {publisher} {John Wiley and Sons},\ \bibinfo {year} {1974})\BibitemShut
  {NoStop}%
\bibitem [{\citenamefont {Okada}\ and\ \citenamefont
  {Yosida}(1973)}]{okada-1973}%
  \BibitemOpen
  \bibfield  {author} {\bibinfo {author} {\bibfnamefont {I.}~\bibnamefont
  {Okada}}\ and\ \bibinfo {author} {\bibfnamefont {K.}~\bibnamefont {Yosida}},\
  }\href {\doibase 10.1143/PTP.49.1483} {\bibfield  {journal} {\bibinfo
  {journal} {Progress of Theoretical Physics}\ }\textbf {\bibinfo {volume}
  {49}},\ \bibinfo {pages} {1483} (\bibinfo {year} {1973})}\BibitemShut
  {NoStop}%
\bibitem [{\citenamefont {Georges}\ \emph {et~al.}(2013)\citenamefont
  {Georges}, \citenamefont {Medici},\ and\ \citenamefont
  {Mravlje}}]{georges-2013}%
  \BibitemOpen
  \bibfield  {author} {\bibinfo {author} {\bibfnamefont {A.}~\bibnamefont
  {Georges}}, \bibinfo {author} {\bibfnamefont {L.~d.}\ \bibnamefont {Medici}},
  \ and\ \bibinfo {author} {\bibfnamefont {J.}~\bibnamefont {Mravlje}},\ }\href
  {\doibase 10.1146/annurev-conmatphys-020911-125045} {\bibfield  {journal}
  {\bibinfo  {journal} {Annual Review of Condensed Matter Physics}\ }\textbf
  {\bibinfo {volume} {4}},\ \bibinfo {pages} {137} (\bibinfo {year}
  {2013})}\BibitemShut {NoStop}%
\bibitem [{\citenamefont {Paredes}\ \emph {et~al.}(2005)\citenamefont
  {Paredes}, \citenamefont {Tejedor},\ and\ \citenamefont
  {Cirac}}]{paredes-2005}%
  \BibitemOpen
  \bibfield  {author} {\bibinfo {author} {\bibfnamefont {B.}~\bibnamefont
  {Paredes}}, \bibinfo {author} {\bibfnamefont {C.}~\bibnamefont {Tejedor}}, \
  and\ \bibinfo {author} {\bibfnamefont {J.~I.}\ \bibnamefont {Cirac}},\ }\href
  {\doibase 10.1103/PhysRevA.71.063608} {\bibfield  {journal} {\bibinfo
  {journal} {Phys. Rev. A}\ }\textbf {\bibinfo {volume} {71}},\ \bibinfo
  {pages} {063608} (\bibinfo {year} {2005})}\BibitemShut {NoStop}%
\bibitem [{\citenamefont {Gorshkov}\ \emph {et~al.}(2010)\citenamefont
  {Gorshkov}, \citenamefont {Hermele}, \citenamefont {Gurarie}, \citenamefont
  {Xu}, \citenamefont {Julienne}, \citenamefont {Ye}, \citenamefont {Zoller},
  \citenamefont {Demler}, \citenamefont {Lukin},\ and\ \citenamefont
  {Rey}}]{gorshkov-2010}%
  \BibitemOpen
  \bibfield  {author} {\bibinfo {author} {\bibfnamefont {A.~V.}\ \bibnamefont
  {Gorshkov}}, \bibinfo {author} {\bibfnamefont {M.}~\bibnamefont {Hermele}},
  \bibinfo {author} {\bibfnamefont {V.}~\bibnamefont {Gurarie}}, \bibinfo
  {author} {\bibfnamefont {C.}~\bibnamefont {Xu}}, \bibinfo {author}
  {\bibfnamefont {P.~S.}\ \bibnamefont {Julienne}}, \bibinfo {author}
  {\bibfnamefont {J.}~\bibnamefont {Ye}}, \bibinfo {author} {\bibfnamefont
  {P.}~\bibnamefont {Zoller}}, \bibinfo {author} {\bibfnamefont
  {E.}~\bibnamefont {Demler}}, \bibinfo {author} {\bibfnamefont {M.~D.}\
  \bibnamefont {Lukin}}, \ and\ \bibinfo {author} {\bibfnamefont {A.~M.}\
  \bibnamefont {Rey}},\ }\href {\doibase 10.1038/nphys1535} {\bibfield
  {journal} {\bibinfo  {journal} {Nat. Phys.}\ }\textbf {\bibinfo {volume}
  {6}},\ \bibinfo {pages} {289} (\bibinfo {year} {2010})}\BibitemShut {NoStop}%
\bibitem [{\citenamefont {Parmenter}(1973)}]{parmenter-1973}%
  \BibitemOpen
  \bibfield  {author} {\bibinfo {author} {\bibfnamefont {R.~H.}\ \bibnamefont
  {Parmenter}},\ }\href {\doibase 10.1103/PhysRevB.8.1273} {\bibfield
  {journal} {\bibinfo  {journal} {Phys. Rev. B}\ }\textbf {\bibinfo {volume}
  {8}},\ \bibinfo {pages} {1273} (\bibinfo {year} {1973})}\BibitemShut
  {NoStop}%
\bibitem [{\citenamefont {M\"uhlschlegel}(1968)}]{muhlschlegel-1968}%
  \BibitemOpen
  \bibfield  {author} {\bibinfo {author} {\bibfnamefont {B.}~\bibnamefont
  {M\"uhlschlegel}},\ }\href {\doibase 10.1007/BF01325759} {\bibfield
  {journal} {\bibinfo  {journal} {Zeitschrift f\"ur Physik}\ }\textbf {\bibinfo
  {volume} {208}},\ \bibinfo {pages} {94} (\bibinfo {year} {1968})}\BibitemShut
  {NoStop}%
\bibitem [{sym()}]{symmetric-KLM-justification}%
  \BibitemOpen
  \href@noop {} {\bibinfo  {journal} {The two-orbital Anderson impurity model
  with $v_{\bp1}\neq v_{\bp2}$ was considered in Refs.
  \cite{eto-2000,pustilnik-2000,kikoin-2001,eto-2001} where it was shown that
  the resulting Kondo Hamiltonian has the same form as our Eq.
  \eqref{so4_KLM}}\ }\BibitemShut {NoStop}%
\bibitem [{GSW()}]{GSWT-finite-J-corrections}%
  \BibitemOpen
\bibfield  {journal} {  }\href@noop {} {\bibinfo  {journal} {Inclusion of the
  Hund splitting of the $N_f=2$ manifold in Eq. \eqref{GSWT} leads to a
  renormalization of $J$ in Eq. \eqref{so4_KLM} by a factor $1+{\rm
  const.}\times(J^2/U^2)+\ldots$}\ }\BibitemShut {NoStop}%
\bibitem [{\citenamefont {Kiselev}(2006)}]{kiselev-2006}%
  \BibitemOpen
\bibfield  {journal} {  }\bibfield  {author} {\bibinfo {author} {\bibfnamefont
  {M.~N.}\ \bibnamefont {Kiselev}},\ }\href {\doibase
  10.1142/S0217979206033310} {\bibfield  {journal} {\bibinfo  {journal}
  {International Journal of Modern Physics B}\ }\textbf {\bibinfo {volume}
  {20}},\ \bibinfo {pages} {381} (\bibinfo {year} {2006})}\BibitemShut
  {NoStop}%
\bibitem [{\citenamefont {Kikoin}\ and\ \citenamefont
  {Avishai}(2001)}]{kikoin-2001}%
  \BibitemOpen
  \bibfield  {author} {\bibinfo {author} {\bibfnamefont {K.}~\bibnamefont
  {Kikoin}}\ and\ \bibinfo {author} {\bibfnamefont {Y.}~\bibnamefont
  {Avishai}},\ }\href {\doibase 10.1103/PhysRevLett.86.2090} {\bibfield
  {journal} {\bibinfo  {journal} {Phys. Rev. Lett.}\ }\textbf {\bibinfo
  {volume} {86}},\ \bibinfo {pages} {2090} (\bibinfo {year}
  {2001})}\BibitemShut {NoStop}%
\bibitem [{\citenamefont {Coqblin}\ and\ \citenamefont
  {Schrieffer}(1969)}]{coqblin-1969}%
  \BibitemOpen
  \bibfield  {author} {\bibinfo {author} {\bibfnamefont {B.}~\bibnamefont
  {Coqblin}}\ and\ \bibinfo {author} {\bibfnamefont {J.~R.}\ \bibnamefont
  {Schrieffer}},\ }\href {\doibase 10.1103/PhysRev.185.847} {\bibfield
  {journal} {\bibinfo  {journal} {Phys. Rev.}\ }\textbf {\bibinfo {volume}
  {185}},\ \bibinfo {pages} {847} (\bibinfo {year} {1969})}\BibitemShut
  {NoStop}%
\bibitem [{\citenamefont {Eto}\ and\ \citenamefont {Nazarov}(2000)}]{eto-2000}%
  \BibitemOpen
  \bibfield  {author} {\bibinfo {author} {\bibfnamefont {M.}~\bibnamefont
  {Eto}}\ and\ \bibinfo {author} {\bibfnamefont {Y.~V.}\ \bibnamefont
  {Nazarov}},\ }\href {\doibase 10.1103/PhysRevLett.85.1306} {\bibfield
  {journal} {\bibinfo  {journal} {Phys. Rev. Lett.}\ }\textbf {\bibinfo
  {volume} {85}},\ \bibinfo {pages} {1306} (\bibinfo {year}
  {2000})}\BibitemShut {NoStop}%
\bibitem [{\citenamefont {Pustilnik}\ and\ \citenamefont
  {Glazman}(2000)}]{pustilnik-2000}%
  \BibitemOpen
  \bibfield  {author} {\bibinfo {author} {\bibfnamefont {M.}~\bibnamefont
  {Pustilnik}}\ and\ \bibinfo {author} {\bibfnamefont {L.~I.}\ \bibnamefont
  {Glazman}},\ }\href {\doibase 10.1103/PhysRevLett.85.2993} {\bibfield
  {journal} {\bibinfo  {journal} {Phys. Rev. Lett.}\ }\textbf {\bibinfo
  {volume} {85}},\ \bibinfo {pages} {2993} (\bibinfo {year}
  {2000})}\BibitemShut {NoStop}%
\bibitem [{\citenamefont {Eto}\ and\ \citenamefont {Nazarov}(2001)}]{eto-2001}%
  \BibitemOpen
  \bibfield  {author} {\bibinfo {author} {\bibfnamefont {M.}~\bibnamefont
  {Eto}}\ and\ \bibinfo {author} {\bibfnamefont {Y.~V.}\ \bibnamefont
  {Nazarov}},\ }\href {\doibase 10.1103/PhysRevB.64.085322} {\bibfield
  {journal} {\bibinfo  {journal} {Phys. Rev. B}\ }\textbf {\bibinfo {volume}
  {64}},\ \bibinfo {pages} {085322} (\bibinfo {year} {2001})}\BibitemShut
  {NoStop}%
\bibitem [{\citenamefont {Anderson}\ and\ \citenamefont
  {Hasegawa}(1955)}]{anderson-1955}%
  \BibitemOpen
  \bibfield  {author} {\bibinfo {author} {\bibfnamefont {P.~W.}\ \bibnamefont
  {Anderson}}\ and\ \bibinfo {author} {\bibfnamefont {H.}~\bibnamefont
  {Hasegawa}},\ }\href {\doibase 10.1103/PhysRev.100.675} {\bibfield  {journal}
  {\bibinfo  {journal} {Phys. Rev.}\ }\textbf {\bibinfo {volume} {100}},\
  \bibinfo {pages} {675} (\bibinfo {year} {1955})}\BibitemShut {NoStop}%
\bibitem [{\citenamefont {Hamada}\ and\ \citenamefont
  {Shimahara}(1995)}]{hamada-1995}%
  \BibitemOpen
  \bibfield  {author} {\bibinfo {author} {\bibfnamefont {M.}~\bibnamefont
  {Hamada}}\ and\ \bibinfo {author} {\bibfnamefont {H.}~\bibnamefont
  {Shimahara}},\ }\href {\doibase 10.1103/PhysRevB.51.3027} {\bibfield
  {journal} {\bibinfo  {journal} {Phys. Rev. B}\ }\textbf {\bibinfo {volume}
  {51}},\ \bibinfo {pages} {3027} (\bibinfo {year} {1995})}\BibitemShut
  {NoStop}%
\bibitem [{\citenamefont {Nagaoka}(1966)}]{nagaoka-1966}%
  \BibitemOpen
  \bibfield  {author} {\bibinfo {author} {\bibfnamefont {Y.}~\bibnamefont
  {Nagaoka}},\ }\href {\doibase 10.1103/PhysRev.147.392} {\bibfield  {journal}
  {\bibinfo  {journal} {Phys. Rev.}\ }\textbf {\bibinfo {volume} {147}},\
  \bibinfo {pages} {392} (\bibinfo {year} {1966})}\BibitemShut {NoStop}%
\bibitem [{\citenamefont {Cox}(1987)}]{cox-1987}%
  \BibitemOpen
  \bibfield  {author} {\bibinfo {author} {\bibfnamefont {D.~L.}\ \bibnamefont
  {Cox}},\ }\href {\doibase 10.1103/PhysRevLett.59.1240} {\bibfield  {journal}
  {\bibinfo  {journal} {Phys. Rev. Lett.}\ }\textbf {\bibinfo {volume} {59}},\
  \bibinfo {pages} {1240} (\bibinfo {year} {1987})}\BibitemShut {NoStop}%
\bibitem [{\citenamefont {Perkins}\ \emph {et~al.}(2007)\citenamefont
  {Perkins}, \citenamefont {N\'u\~nez Regueiro}, \citenamefont {Coqblin},\ and\
  \citenamefont {Iglesias}}]{perkins-2007}%
  \BibitemOpen
  \bibfield  {author} {\bibinfo {author} {\bibfnamefont {N.~B.}\ \bibnamefont
  {Perkins}}, \bibinfo {author} {\bibfnamefont {M.~D.}\ \bibnamefont {N\'u\~nez
  Regueiro}}, \bibinfo {author} {\bibfnamefont {B.}~\bibnamefont {Coqblin}}, \
  and\ \bibinfo {author} {\bibfnamefont {J.~R.}\ \bibnamefont {Iglesias}},\
  }\href {\doibase 10.1103/PhysRevB.76.125101} {\bibfield  {journal} {\bibinfo
  {journal} {Phys. Rev. B}\ }\textbf {\bibinfo {volume} {76}},\ \bibinfo
  {pages} {125101} (\bibinfo {year} {2007})}\BibitemShut {NoStop}%
\bibitem [{\citenamefont {Thomas}\ \emph {et~al.}(2011)\citenamefont {Thomas},
  \citenamefont {da~Rosa Sim\~oes}, \citenamefont {Iglesias}, \citenamefont
  {Lacroix}, \citenamefont {Perkins},\ and\ \citenamefont
  {Coqblin}}]{thomas-2011}%
  \BibitemOpen
  \bibfield  {author} {\bibinfo {author} {\bibfnamefont {C.}~\bibnamefont
  {Thomas}}, \bibinfo {author} {\bibfnamefont {A.~S.}\ \bibnamefont {da~Rosa
  Sim\~oes}}, \bibinfo {author} {\bibfnamefont {J.~R.}\ \bibnamefont
  {Iglesias}}, \bibinfo {author} {\bibfnamefont {C.}~\bibnamefont {Lacroix}},
  \bibinfo {author} {\bibfnamefont {N.~B.}\ \bibnamefont {Perkins}}, \ and\
  \bibinfo {author} {\bibfnamefont {B.}~\bibnamefont {Coqblin}},\ }\href
  {\doibase 10.1103/PhysRevB.83.014415} {\bibfield  {journal} {\bibinfo
  {journal} {Phys. Rev. B}\ }\textbf {\bibinfo {volume} {83}},\ \bibinfo
  {pages} {014415} (\bibinfo {year} {2011})}\BibitemShut {NoStop}%
\bibitem [{\citenamefont {Diep}(2004)}]{diep-2004}%
  \BibitemOpen
  \bibfield  {author} {\bibinfo {author} {\bibfnamefont {H.}~\bibnamefont
  {Diep}},\ }\href {http://books.google.com/books?id=eVZmjOvkelUC} {\emph
  {\bibinfo {title} {Frustrated Spin Systems}}}\ (\bibinfo  {publisher} {World
  Scientific Publishing Company, Incorporated},\ \bibinfo {year}
  {2004})\BibitemShut {NoStop}%
\end{thebibliography}
\end{document}